\documentclass[twocolumn,english]{revtex4-1}
\usepackage[T1]{fontenc}
\usepackage[latin9]{inputenc}
\usepackage{color}
\usepackage{amsmath}
\usepackage{graphicx}
\usepackage{esint}

\makeatletter

\usepackage{xfrac}

\usepackage{babel}

\usepackage{babel}

\makeatother

\usepackage{babel}
\begin{document}
\title{Stability of the smectic phase in arrays of parallel quantum wires}
\author{Geo Jose and Bruno Uchoa}
\affiliation{Center for Quantum Research and Technology, The University of Oklahoma,
Norman, Oklahoma, 73071, USA}
\begin{abstract}
Using bosonization, we study a model of parallel quantum wires constructed
from two dimensional Dirac fermions in the presence of periodic topological
domain walls. The model accounts for the lateral spread of the wavefunctions
$\ell$ in the transverse direction to the wires. The gapless modes
confined to each domain wall are shown to form Luttinger liquids,
which realize a well known smectic non-Fermi liquid fixed point when
interwire Coulomb interactions are taken into account. Perturbative
studies on phenomenological models have shown that the smectic fixed
point is unstable towards a variety of phases such as superconductivity,
stripe, smectic and Fermi liquid phases. Here, we show that the considered
model leads to a phase diagram with only smectic metal and Fermi liquid
phases. The smectic metal phase is stable in the ideal quantum wire
limit $\ell\to0$. For finite $\ell$, we find a critical Coulomb
coupling separating the strong coupling smectic metal from a weak
coupling Fermi liquid phase. We conjecture that the absence of superconductivity
should be a generic feature of similar models. \textcolor{black}{We
discuss the relevance of this model for quantum wires created with
moire heterostructures.}
\end{abstract}
\maketitle

\section{\textit{\emph{Introduction}}}

The low energy properties of gapless, interacting fermions in one
dimension are generically described by the Luttinger liquid (LL) theory.
The low energy excitations in this universality class are density
waves, unlike quasiparticles in Fermi liquid theory in two or three
dimensions \cite{Giamarchi,Fradkin}. Arrays of parallel LL's have
been studied extensively in the past thirty years with motivations
such as the possibility of constructing higher dimensional non Fermi
liquids and understanding the unusual normal state in cuprate superconductors
\cite{Strong,Anderson,Tranquanda,Zannen,Machida,Kato,Emery-3,Mukhopadhyay-1}.
Perturbations to the decoupled parallel LL arrays were generically
shown to result in a higher dimensional Fermi liquid or an ordered
state \cite{Emery-1,Bourbonnais}. It was then pointed out that one
needs to include other marginal operators such as inter-wire density-density
and current-current interactions in the most general fixed point action,
which then describes a generalized smectic state \cite{Emery}. The
perturbative stability of the smectic fixed is now understood and
leads to a rich, if generic, phase diagram, where the following phases
can arise: \emph{i) }smectic superconductor, \emph{ii) }insulating
stripe crystal, \emph{iii) }Fermi liquid and \emph{iv) }smectic metal
state \cite{Emery,Vishwanath,Mukhopadhyay}. Very recently, this theory
has been applied to the case of a triangular network of LL's \cite{Chen}
that emerge naturally in twisted bilayer graphene at marginal angles
\cite{Xu}. However, all these works are phenomenological. In this
paper, we study an \textcolor{black}{effective} model within the formalism
established by the hitherto mentioned works and examine what part
of the generic phase diagram actually survives.

\begin{figure}[t]
\begin{centering}
\includegraphics[scale=0.3]{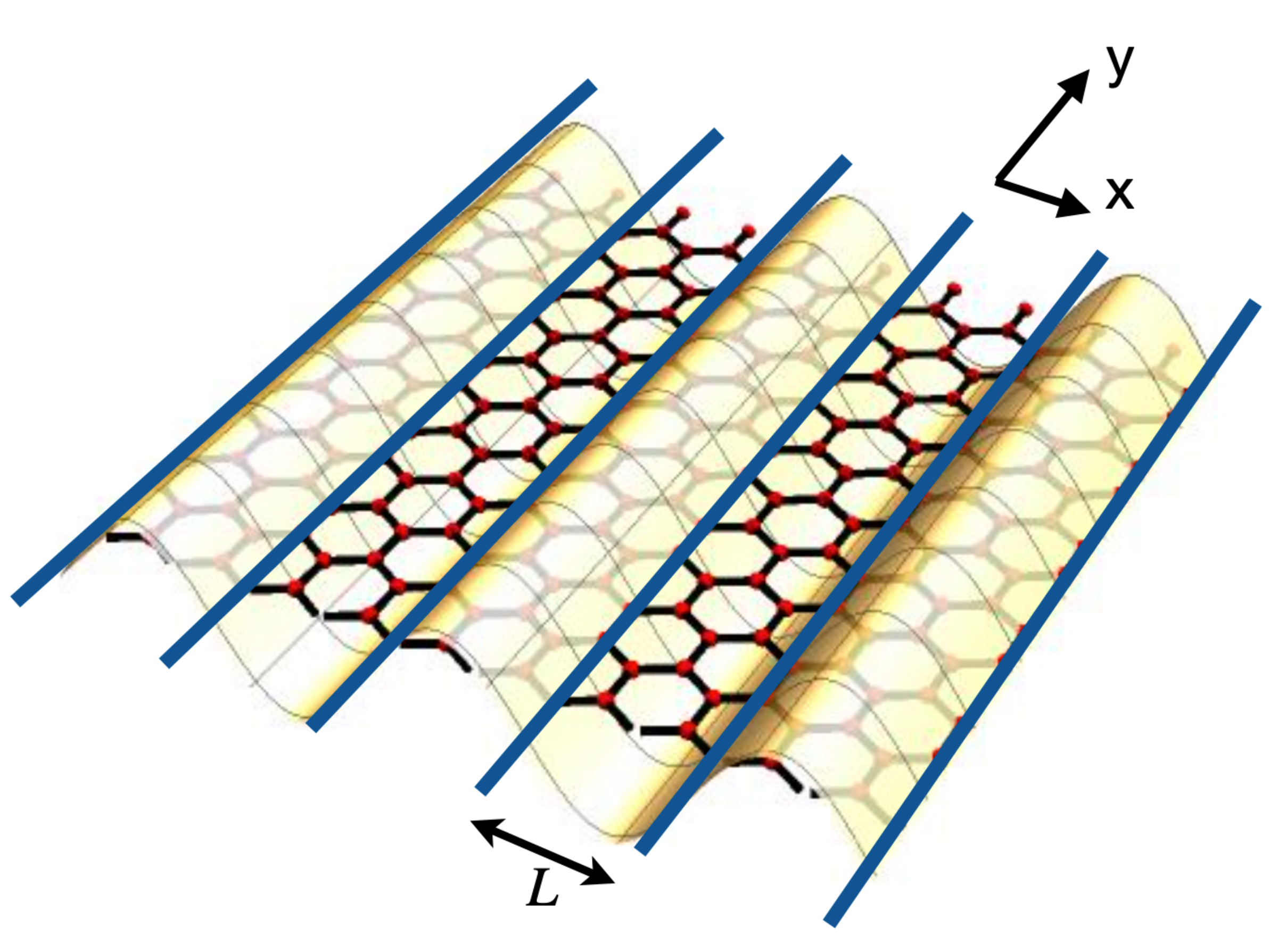}
\par\end{centering}
\caption{Cartoon representation of a periodic mass term (yellow wave form)
on the honeycomb lattice. One dimensional modes at interfaces where
the mass term changes sign are shown as thick blue lines. The transverse
distance between these modes is $L$. The modes that live on the interface
have a transverse width that dies off as $\text{e}^{-x^{2}/(2\ell^{2})}$,
with $\ell$ the lateral spread of the wavefunction (see text).}
\end{figure}

\textcolor{black}{We first point out that moire heterostructures \cite{Andrei}
offer a formidable platform for building arrays of identical and perfectly
spaced quantum wires. The latter were recently realized in moire superlattices
of twisted bilayer tungsten ditelluride (tWTe$_{2}$), which display
parallel LL transport above $1.8$K \cite{Wang}. The distance between
quantum wires is tunable by the interlayer twist angle. In general,
application of a voltage difference between two layers in a van der
Waals heterostructure (achievable by separately controlling the gates
at each layer) permits theoretically integrating out the degrees of
freedom in one of the layers. This results in an effective description
for the other layer, that is affected by local potentials that modulate
with the period of the moire pattern \cite{Kindermann-1}. For tWTe$_{2}$,
those local potentials make the bands flat along one direction, confining
electrons along quasi-1D modes. In the case of moire heterostructures
with Dirac fermions in each monolayer, as in marginally twisted graphene
bilayers, those local potentials create periodic scalar, vector and
mass terms \cite{Uchoa,Yankowitz,Yang,Sachs}. A periodic mass term
can confine Dirac fermions, with the lines where it changes sign forming
quasi-1D modes that are topological. The periodic vector potential
can be gauged away \cite{Kindermann}, and the scalar potential controlled
with gating effects.}

We construct an effective model for an array of parallel quantum wires
using 2D Dirac fermions in the presence of a periodically modulated
mass term. This potential confines the low energy quasiparticles to
propagate along one dimensional channels, as depicted in Fig. 1. Incorporating
intra-wire Coulomb interactions, these modes are shown to be tunable
LL's, akin to the domain wall modes found in gated bilayer graphene
\cite{Killi,Martin}, in mono and bilayer graphene under irradiation
\cite{Biswas}, and also in other contexts \cite{Heeger,Semenoff,Kindermann}.
When inter-wire Coulomb interactions are taken into account, the model
realizes a smectic metal state, whose Luttinger parameters can be
obtained in terms of the fine structure constant of the material and
the lateral confinement of the wavefunctions near the wires.

\textcolor{black}{Even though the proposed model is topological, we
suggest that it offers insight on the stability of the smectic phase
for parallel quantum wires with non-topological origin, as in tWTe$_{2}$,
and arrays of generic quantum wires with exponentially localized wavefunctions.}\textcolor{blue}{{}
}This model naturally incorporates the lateral spread of the wavefunctions
$\ell$ in the perpendicular direction to the quantum wires. Employing
well established abelian bosonization methods to account for the possible
instabilities to the smectic fixed point, we show that only two phases
remain in this model: a smectic metal state and a 2D Fermi liquid
state. We point out that, while the absence of the insulating smectic
(stripe) phase is due to the lack of backscattering in the model,
superconductivity is shown to be always an irrelevant (or marginal)
perturbation at the smectic fixed point.

In the phase diagram, we find that the smectic metal and the Fermi
liquid phases are separated by a quantum critical point set by a critical
Coulomb coupling $\alpha_{c}$, whose value is determined by the lateral
spread $\ell$ normalized by the distance between the wires. For finite
$\ell$, the Fermi liquid phase is the leading instability in weak
coupling $(\alpha<\alpha_{c}$), while the smectic metal is dominant
in the strong coupling regime $(\alpha>\alpha_{c})$. The critical
coupling vanishes in the ideal quantum wire limit $(\ell\to0$), where
the smectic phase is always dominant. \textcolor{black}{We finally
discuss the role of weak backscattering effects in similar models
and suggest that the absence of superconductivity should be a general
feature of quantum wires in moire heterostructures.}

The paper is organized in the following way: in section II we describe
the model of parallel quantum wires and derive the wavefunctions of
their low energy modes. We then address the bosonized action of the
quantum wires as tunnable LLs, with the Luttinger parameters expressed
in terms of the fine structure constant $\alpha$ and the lateral
spread $\ell$. We then write the action at the smectic metal fixed
point once inter-wire Coulomb interactions are accounted for. In section
III, we derive the phase diagram with the possible instabilities of
the smectic fixed point according to this model, with a discussion
about backscattering effects. Finally, in section IV we present our
conclusions.

\section{Model}

We consider a generic continuum model of 2D Dirac fermions with two
valley flavors, as in the honeycomb lattice. Generalizations to other
physical lattices with Dirac quasiparticles and an arbitrary number
of valleys are straightforward. In real space, the free Hamiltonian
is
\begin{equation}
\mathcal{H}_{0}=\int\text{d}^{2}r\,\sum_{\sigma}\Psi_{\sigma}^{\dagger}(\mathbf{r})\hat{\mathcal{H}}_{0}(\mathbf{r})\Psi_{\sigma}(\mathbf{r}),\label{eq:Ho}
\end{equation}
where
\begin{equation}
\hat{\mathcal{H}}_{0}(\mathbf{r})=\left(\begin{array}{cc}
\hat{\mathcal{H}}_{+} & 0\\
0 & \hat{\mathcal{H}}_{-}
\end{array}\right)\label{eq:Ho0}
\end{equation}
is a $4\times4$ matrix defined in the two valleys $\alpha=\pm$,
\begin{equation}
\mathcal{H}_{\nu}(\mathbf{r})=-iv\sigma_{x}\partial_{x}-i\alpha v\sigma_{y}\partial_{y}+M(\mathbf{r})\sigma_{z},\label{eq:Ho1}
\end{equation}
with $\sigma_{x}$ and $\sigma_{y}$ the off diagonal Pauli matrices
in the pseudospin space. $\Psi_{\sigma}(\mathbf{r})$ is a four-component
spinor with spin $\sigma=\uparrow,\downarrow.$ The mass term profile
is taken to be of the form
\begin{equation}
M(\mathbf{r})=M_{0}\sin\left(\frac{\pi x}{L}\right),\label{Mass term}
\end{equation}
which breaks the continuous translational symmetry along the $x$
direction. It is well known in the context of the index theorem that
real space lines where the mass term changes sign are topological,
hosting zero energy modes \cite{Jackiw}. The mass term \eqref{Mass term}
is a periodic function that changes sign at the nodal lines where
$M(\mathbf{r})=0$, forming an array of parallel quantum wires with
spacing $L$ shown in Fig. 1.

Before addressing the fate of the possible many-body phases in this
system, we first compute the zero-modes that live on these nodal lines.
To solve for the eigenvalues and eigenvectors of $\mathcal{H}_{+}(\mathbf{r}),$
it is convenient to linearize Eq. \eqref{Mass term} in the vicinity
of a zero-mass line at $x=0$ to get
\begin{equation}
M(\mathbf{r})\approx M_{0}\frac{\pi x}{L}.\label{eq:linearized Mass}
\end{equation}
The eigenvalue problem can be solved analytically by performing two
sequential unitary transformations in the pseudospin \cite{Tchoumakov}:
a rotation by $-\pi/2$ around the $z$ axis, that takes $\sigma_{x}\rightarrow\sigma_{y}$
and $\sigma_{y}\rightarrow-\sigma_{x}$, followed by a rotation by
$\pi/2$ around the $y$ axis, which takes $\sigma_{x}\rightarrow-\sigma_{z}$
and $\sigma_{z}\rightarrow\sigma_{x}.$ In the transformed basis,
the ``$+$'' block of the eigenvalue problem
\begin{equation}
\mathcal{H}(\mathbf{r})\Psi(\mathbf{r})=E\Psi(\mathbf{r})\label{eq:Eigenvalue}
\end{equation}
 can be written as
\begin{equation}
\omega\begin{pmatrix}\ell k_{y} & -\partial_{\xi}+\xi\\
\partial_{\xi}+\xi & -\ell k_{y}
\end{pmatrix}\Phi_{+}(\xi)=E_{+}\Phi_{+}(\xi),\label{dimensionless-H-plus}
\end{equation}
where we have introduced variables
\begin{equation}
\ell=\sqrt{\frac{vL}{M_{0}\pi}},\label{eq:ell}
\end{equation}
$\omega=v/\ell$ and $\xi=x/\ell.$ Eq. (\ref{dimensionless-H-plus})
implicitly assumes the ansatz 
\begin{equation}
\Psi_{+}(\mathbf{r})=\frac{e^{ik_{y}y}}{\sqrt{L_{y}}}\left(\begin{array}{c}
\Phi_{+}(x)\\
\mathbf{0}
\end{array}\right)\label{eq:Psi+}
\end{equation}
 due to translational symmetry in the $y$ direction, with $k_{y}$
the corresponding momentum and $L_{y}$ the length of the quantum
wires.

In this form, Eq. \eqref{dimensionless-H-plus} resembles the problem
of Dirac fermions in the presence of a uniform magnetic field \cite{Castro-Neto},
$\ell$ being the analogue of the magnetic length. Defining the ladder
operator $\mathcal{O}=\left(\partial_{\xi}+\xi\right)/\sqrt{2}$ such
that $\left[\mathcal{O},\mathcal{O}^{\dagger}\right]=1$ and the number
operator $\hat{N}=\mathcal{O}^{\dagger}\mathcal{O},$ one can easily
infer that the eigenvalues are given by
\begin{equation}
E_{+,N}^{(\pm)}(k_{y})=\pm\omega\sqrt{\ell^{2}k_{y}^{2}+2N},\label{eq:E}
\end{equation}
 where $N=1,2,\ldots$ indexing the gapped quantum wire modes, with
the corresponding eigenvectors
\begin{equation}
\Phi_{+}^{N,\pm}(\xi)=\begin{pmatrix}\psi_{N}(x)\\
\pm\psi_{N-1}(x)
\end{pmatrix}.\label{eq:Phi1}
\end{equation}
In a more explicit form, 
\begin{equation}
\psi_{N}(x)=\frac{2^{-\frac{N}{2}}}{\pi^{\frac{1}{4}}\sqrt{N!}}\text{e}^{-x^{2}/(2\ell^{2})}H_{N}(\xi),\label{eq:HN}
\end{equation}
where $H_{N}(\xi)$ is the $N$-th Hermite polynomial. The length
$\ell$ hence determines the lateral spread of the wavefunctions confined
to the quantum wires. The solution of the $\Psi_{-}(\mathbf{r})$
eigenmodes in the opposite valley is related by time-reversal operation,
\begin{equation}
\Psi_{-,N}^{(\pm)}(\mathbf{r})=\frac{e^{-ik_{y}y}}{\sqrt{L_{y}}}\left(\begin{array}{c}
\mathbf{0}\\
\Phi_{-}^{N,\pm}(\xi)
\end{array}\right),\label{eq:Psi -}
\end{equation}
with $\Phi_{-}^{N,\pm}(\xi)=\left(\psi_{N}(x),\pm\psi_{N-1}(x)\right)^{T}$.
The ``$\pm$'' upper index accounts for the two particle-hole branches
in each valley.

The $N=0$ case corresponds to the gapless zero energy modes moving
along the quantum wires. This case requires a more careful analysis
to resolve the seeming ambiguity between particle and hole states
in Eq. (\ref{eq:E}). In valley $\alpha=+$, the $x$ dependent part
of the wavefunction 
\begin{equation}
\Psi_{+}(x)=\left(\begin{array}{c}
\Phi_{+}^{0}(\xi)\\
\mathbf{0}
\end{array}\right)\label{eq:Psi+zero}
\end{equation}
gives the four-component eigenvector for a single right moving mode
(per spin) with energy dispersion $E_{+}(k_{y})=vk_{y}$. That can
be seen by plugging in $\Phi_{+}^{0}(\xi)$ into \eqref{dimensionless-H-plus}
and explicitly solving the resulting differential equation. The zero
energy mode $\Psi_{-}(x)=\left(\mathbf{0},\Phi_{-}^{0}(\xi)\right)^{T}$in
the opposite valley corresponds to a left moving mode with energy
$E_{-}(k_{y})=-vk_{y}$, as required by time reversal symmetry.

\subsection{\textit{\emph{Tunable LLs}} }

To derive an effective one dimensional model, we assume a suitable
energy cut off $v\Lambda$ below the bulk mass $M_{0}$ and focus
on the gapless modes propagating along the quantum wires. We closely
follow the LL derivation in Ref. \cite{Killi,Biswas}. In the infrared,
we restrict our interest to the gapless modes with $N=0$, with $k_{y}$
the small momentum in the vicinity of the two valleys, $\alpha K$.
The field operator becomes 
\begin{equation}
\hat{\chi}_{\sigma}(\mathbf{r})=\frac{1}{\sqrt{L_{y}}}\sum_{\alpha=\pm}e^{i\alpha Ky}\Psi_{\alpha}(x)\hat{\zeta}_{\alpha,\sigma}(y)\label{Field operator}
\end{equation}
where $\hat{\zeta}_{\alpha,\sigma}(y)=\sum_{k_{y}}\text{e}^{ik_{y}y}\hat{\zeta}_{\alpha,\sigma,k_{y}}$
is a slowly varying field operator for electrons in valley $\alpha$
moving along the wire. The non-interacting Hamiltonian is given by
\begin{equation}
\mathcal{H}_{0}=v\sum_{k_{y},\sigma,\alpha}\alpha k_{y}\hat{\zeta}_{\alpha,\sigma}^{\dagger}(k_{y})\hat{\zeta}_{\alpha,\sigma}(k_{y}).\label{eq:H0}
\end{equation}
The effective Coulomb interaction projected onto the one dimensional
modes can be obtained by substituting \eqref{Field operator} into
the Coulomb interaction term
\begin{equation}
\mathcal{H}_{I,\text{intra}}=\frac{1}{2}\sum_{\sigma,\sigma'}\int_{\mathbf{r,r'}}\hat{\rho}(\mathbf{r})V(\mathbf{r}-\mathbf{r'})\hat{\rho}(\mathbf{r}^{\prime}),\label{eq:Ho-1}
\end{equation}
defined in terms of density operators
\begin{equation}
\hat{\rho}(\mathbf{r})=\sum_{\sigma}\hat{\chi}_{\sigma}^{\dagger}(\mathbf{r})\hat{\chi}_{\sigma}(\mathbf{r}).\label{eq:rho}
\end{equation}
Here, $V(\mathbf{r})=e^{2}\text{e}^{-r/\lambda}/(\epsilon r)$ is
a screened Coulomb interaction, with the screening length $\lambda$
set by metallic contacts with the wires and $\epsilon$ the background
dielectric constant.

We note that the orthogonality of the spinors $\Psi_{+}(x)$ and $\Psi_{-}(x)$
suppresses backscattering in this model, unlike in conventional LLs.
Since $\hat{\zeta}_{\alpha,\sigma}(y)$ are slow varying fields, the
effective intra-wire interaction can be approximated by
\begin{equation}
\mathcal{H}_{I,\text{intra}}\approx\int_{y}\sum_{\alpha\beta}g_{\alpha\beta}\hat{\zeta}_{\alpha,\sigma}^{\dagger}(y)\hat{\zeta}_{\beta,\sigma'}^{\dagger}(y)\hat{\zeta}_{\beta,\sigma'}(y)\hat{\zeta}_{\alpha,\sigma}(y),\label{eq:H Intra}
\end{equation}
with
\begin{equation}
g_{\alpha\beta}=\frac{1}{2}\int_{x,x'}\int_{\bar{y}}V(x-x',\bar{y})\left|\psi_{0}\left(x\right)\right|^{2}\left|\psi_{0}\left(x^{\prime}\right)\right|^{2},\label{eq:g}
\end{equation}
where $\bar{y}=y-y^{\prime}$. Using the standard g-ology notation
in the LL literature, the denote $g_{+-}=g_{-+}=g_{2}$ and $g_{--}=g_{++}=g_{4}$,
which turn out to be the same, $g_{2}=g_{4}$. This is not a coincidence,
but a manifestation of the chiral symmetry of the problem in the forward
scattering terms. The equality between $g_{2}$ and $g_{4}$ also
implies in the absence of current-current interaction terms \cite{note0}.

In order to bosonize the fermionic Hamiltonian, we follow the abelian
bosonization convention in Ref. \cite{Fradkin,Emery}. The fermionic
fields for left and right moving modes
\begin{equation}
\hat{\zeta}_{\alpha,\sigma}(y)\sim e^{i\alpha\sqrt{\pi}\left[\phi_{\sigma}(y)-\alpha\theta_{\sigma}(y)\right]}\label{eq:zeta}
\end{equation}
are cast in terms of the two bosonic fields $\phi_{\sigma}(y)$ and
$\theta_{\sigma}(y)$. The Hamiltonian $\mathcal{H}_{\mathrm{intra}}=\mathcal{H}_{0}+\mathcal{H}_{I,\mathrm{intra}}$
written in terms of charge $(\rho)$ and spin $(\sigma)$ variables
exhibit spin-charge separation, 
\begin{equation}
\mathcal{H}_{\rho,\sigma}=\frac{1}{2}\int dy\left[\left(\partial_{y}\Theta_{\rho,\sigma}\right)^{2}\frac{u_{\rho,\sigma}}{K_{\rho,\sigma}}+\left(\partial_{y}\Phi_{\rho,\sigma}\right)^{2}u_{\rho,\sigma}K_{\rho,\sigma}\right],\label{Hrhosigma}
\end{equation}
where 
\begin{equation}
\Theta_{\rho,\sigma}(y)=\frac{1}{\sqrt{2}}[\theta^{\uparrow}(y)\pm\theta^{\downarrow}(y)]\label{eq:Theta}
\end{equation}
 and
\begin{equation}
\Phi_{\rho,\sigma}(y)=\frac{1}{\sqrt{2}}[\phi^{\uparrow}(y)\pm\phi^{\downarrow}(y)].\label{eq:Phi}
\end{equation}
The Luttinger parameters are given by $u_{\rho,\sigma}=v$, $K_{\sigma}=1$
and
\begin{equation}
K_{\rho}=\left[1+2sg_{4}/(\pi v)\right]^{\frac{1}{2}}.\label{eq:Kappa}
\end{equation}
In the above, $s=2$. In the case of spinless fermions, the spin part
of $\mathcal{H}_{\text{intra}}$ is absent and $s=1$. The LL stiffness
in the charge sector $K_{\rho}$ can be controlled by tuning the lateral
spread of the wavefunctions $\ell$ through the coupling $g_{4}$.
The one dimensional modes that live on the nodal lines thus form a
lattice of decoupled LLs.

\subsection{\textit{\emph{Smectic metal}}}

The density-density interaction between wires follows from $\mathcal{H}_{I,\mathrm{intra}}$
after incorporating the wire index $a$ for the superlattice into
the wavefunctions and $\hat{\xi}_{\alpha,\sigma}^{a}(y)$ operators,
and hence into the definition of the field operators, $\hat{\chi}_{\sigma,a}(\mathbf{r})=\sum_{\alpha=\pm}e^{i\alpha Ky}\Psi_{\alpha,\sigma}(x_{a})\hat{\zeta}_{\alpha,\sigma}^{a}(y)$,
with $x_{a}\equiv x-X_{a}$ the relative coordinate to wire $a$.
In line with \cite{Emery,Vishwanath,Mukhopadhyay}, we only consider
the coupling of charge densities between wires and not the exchange
coupling, which is small when $\ell/L\ll1$, with $L$ the interwire
distance. Thus,
\begin{equation}
\mathcal{H}_{\mathrm{inter}}=\frac{1}{2}\sum_{a\neq a'}\int_{\mathbf{r,r'}}V(\mathbf{r-r}^{\prime})\rho_{a}(\mathbf{r})\rho_{a'}(\mathbf{r}^{\prime}).\label{eq:Inter}
\end{equation}
From the bosonization identities, we can write this as an effective
one dimensional density density interaction, 
\begin{equation}
\mathcal{H}_{\mathrm{inter}}=\frac{2}{\pi}\int_{y}\sum_{a\neq a'}U_{a,a'}\left[\partial_{y}\Phi_{\rho,a}(y)\right]\left[\partial_{y}\Phi_{\rho,a'}(y)\right]\label{Bosonized-inter-wire-Hamiltonian}
\end{equation}
where
\begin{equation}
U_{a,a^{\prime}}=\frac{1}{2}\int_{x,x',\bar{y}}V(x-x^{\prime},\bar{y})\left|\psi_{0}(x_{a})\right|^{2}\left|\psi_{0}(x'_{a^{\prime}})\right|^{2}.\label{U}
\end{equation}
 Thus only the charge sector is modified by the interwire interaction. 

The action for the spin and charge degrees of freedom can be obtained
by integrating out the $\Theta^{\rho,\sigma}$ fields. This yields
\begin{eqnarray}
\mathcal{S}^{\sigma} & = & \int_{k,k_{\bot},\omega}\frac{K^{\sigma}}{2}\left(\frac{\omega^{2}}{u_{\sigma}}+u^{\sigma}k^{2}\right)\left|\Phi_{\sigma}(\mathbf{k})\right|^{2}\label{spin-action}\\
\mathcal{S}^{\rho} & = & \int_{k,k_{\bot},\omega}\frac{K^{\rho}(k_{\bot})}{2}\left(\frac{\omega^{2}}{u^{\rho}(k_{\bot})}+u^{\rho}(k_{\bot})k^{2}\right)\left|\Phi_{\rho}(\mathbf{k})\right|^{2}\qquad\label{charge-action}
\end{eqnarray}
where $u^{\sigma}=v$, $K^{\sigma}=1$, as before, and
\begin{equation}
\int_{k,k_{\bot},\omega}\equiv L/(2\pi)^{3}\int_{-\infty}^{\infty}d\omega dk\int_{-\pi/L}^{\pi/L}dk_{\bot}.\label{eq:int}
\end{equation}
 Luttinger parameters of the charge sector acquire momentum dependence
from \eqref{Bosonized-inter-wire-Hamiltonian}. Stability of the theory
requires these parameters to be positive. We restrict the sum in \eqref{Bosonized-inter-wire-Hamiltonian}
to nearest neighbor interactions as in \cite{Emery}. This gives
\begin{eqnarray}
\frac{u^{\rho}(k_{\bot})}{K^{\rho}(k_{\bot})} & = & v\nonumber \\
u^{\rho}(k_{\bot})K^{\rho}(k_{\bot}) & = & v+\frac{1}{\pi}2sg_{4}+\frac{1}{\pi}4sV_{1}\cos k_{\bot}L,\label{Luttinger-parameters-smectic}
\end{eqnarray}
where $V_{1}=U_{a'=a\pm1},$ $\mathbf{k}=\left(\omega,k,k_{\bot}\right),$
$s=2$. In the spinless case, $s=1$ and the spin part of the action
$S^{\sigma}$ is absent \cite{text-1}. This phase is known as a smectic
metal \cite{Emery,Vishwanath,Mukhopadhyay}. 

\section{\textit{\emph{Phase diagram}}}

The stability of the smectic metal state to various instabilities
has to be assessed via a renormalization group (RG) analysis of the
relevant perturbations. Vast literature exist on the RG analysis of
the smectic fixed point \cite{Emery,Vishwanath,Mukhopadhyay}. Therefore
we do not repeat the analysis here, but adapt their RG equations to
our model. The potentially relevant interactions in this case are
nearest neighbor single electron tunneling $(\mathcal{H}_{t})$, nearest
neighbor singlet pair (Josephson) tunneling $(\mathcal{H}_{sc})$,
and the coupling between the charge density wave (CDW) order parameters.
As mentioned before, due to the absence of backscattering, the interaction
between CDW order parameters are absent in our model. The former two
are given by
\begin{eqnarray}
\mathcal{H}_{t} & = & \mathcal{T}\sum_{a,\alpha,\sigma}\int dx\hat{\zeta}_{\alpha,\sigma}^{\dagger a}\hat{\zeta}_{\alpha,\sigma}^{a+1}+h.c.\label{Ht}\\
\mathcal{H}_{sc} & = & \mathcal{J}\sum_{a,\alpha,\alpha'}\int dx\hat{\zeta}_{\alpha,\uparrow}^{\dagger a}\hat{\zeta}_{-\alpha,\downarrow}^{\dagger a}\hat{\zeta}_{\alpha',\downarrow}^{a+1}\hat{\zeta}_{-\alpha',\uparrow}^{a+1}+h.c.\:,\label{Hsc}
\end{eqnarray}
where $\mathcal{T}$ and $\mathcal{J}$ are the single particle and
Josephson tunneling amplitudes, respectively.

These perturbations become relevant when their scaling dimensions,
$\eta_{X}=2-\triangle_{X}>0$ for $X=t,sc.$ These scaling dimensions
obtained from a one loop RG analysis in the spinless case are \cite{text-2},
\begin{eqnarray}
\triangle_{sc} & = & \int_{-\pi}^{\pi}\frac{dk_{\bot}}{2\pi}\left[2\kappa\left(k_{\bot}\right)\right]\left(1-\cos k_{\bot}\right)\\
\triangle_{t} & = & \frac{1}{4}\int_{-\pi}^{\pi}\frac{dk_{\bot}}{2\pi}2\left[\kappa\left(k_{\bot}\right)+\kappa^{-1}\left(k_{\bot}\right)\right]\left(1-\cos k_{\bot}\right),\qquad
\end{eqnarray}
where in the present model
\begin{equation}
\kappa\left(k_{\bot}\right)=\sqrt{\left(1+\frac{2sg_{4}}{\pi v_{F}}+\frac{4sV_{1}}{\pi v_{F}}\cos k_{\bot}\right)}.\label{kappa}
\end{equation}
In the spinful case, $\triangle_{sc}^{\text{spin}}=1+\frac{1}{2}\triangle_{sc}$
and $\triangle_{t}^{\text{spin}}=\frac{1}{2}+\frac{1}{2}\triangle_{t}.$

We plot in Fig. 2 the regions where these perturbations are relevant
as a function of fine structure constant, $\alpha_{f}=e^{2}/(\epsilon v)$
and the dimensionless lateral spread of the wavefunction in the wires,
$\ell/L$. \textcolor{black}{We restrict the analysis to the regime
$\ell/L\ll1$, where the smectic action is stable. There is no part
of the phase diagram where superconductivity is relevant \cite{Text}.
It has been phenomenologically proposed that superconductivity may
result in }\textit{\textcolor{black}{active}}\textcolor{black}{{} environments,
such as in high-$T_{c}$ scenarios \cite{Emery-2}. In the present
model with screened Coulomb interactions, the minimum value of $\triangle_{sc}$
is $2,$ making it marginal at best. A similar conclusion is applicable
to non-topological quantum wires with repulsive interactions whenever
$g_{4}\gg V_{1}$. }

The curves in the plot describe the critical coupling $\alpha_{c}$
separating the regions where the smectic metal and the Fermi liquid
phases emerge. As previously announced, the 2D Fermi liquid phase
is the dominant instability in the weak coupling regime, when $\alpha_{f}<\alpha_{c}(\ell)$,
whereas the smectic metal phase is the most relevant perturbation
in strong coupling, $\alpha_{f}>\alpha_{c}(\ell)$. The quantum critical
phase transition collapses in the $\ell\to0$ limit, where $\alpha_{c}(\ell)$
scales to zero. That limit corresponds to the physical situation where
the amplitude of the mass term in (\ref{Mass term}) $M_{0}\gg vL$.
The solid lines describe the spinless case, when the screening length
$\lambda/L=0.5$ (black triangles) and 1 (purple circles). The other
two dashed curves correspond to the spinfull case for $\lambda/L=0.5$
(blue squares) and $1$ (orange diamonds). 

The quantum wires described by this model remain generically stable
in the regime where Coulomb repulsion (parametrized by the fine structure
constant $\alpha_{f}$) is strong enough, with the critical coupling
$\alpha_{c}$ set by the lateral spread $\ell$ in the transverse
direction to the wires. In the smectic metal phase, repulsive inter-wire
interactions suppress single particle hopping between the wires, keeping
the unidirectional flow of electrons stable. Conversely, when interactions
are sufficiently weak ($\alpha_{f}<\alpha_{c}$), single particle
tunneling becomes a relevant perturbation to the smectic metal, driving
electrons to percolate accross the quantum wires for any finite $\ell$.
In the latter regime, coherent unidirectional transport along the
wires is destroyed in favor an isotropic 2D Fermi liquid. 

\textcolor{black}{The proposed quantum phase transition can be experimenatlly
explored through the control of the background dielectric constant
of the substrate in the heterostructure where the wires are constructed
\cite{Kim2}. We predict that dielectric materials could be suitably
employed to push the system accross the quantum phase transition. }

\begin{figure}[t]
\begin{centering}
\includegraphics[scale=0.38]{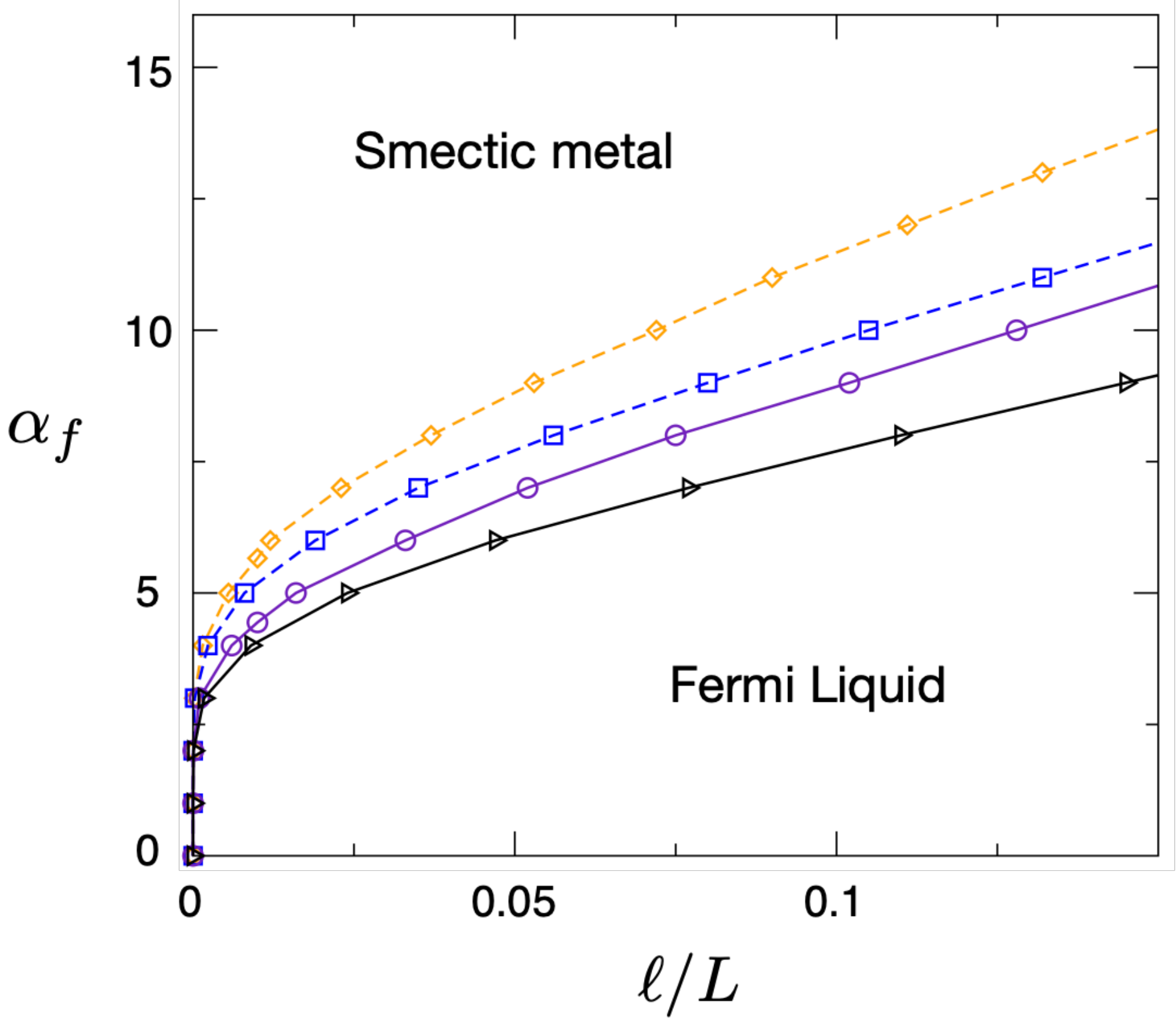}
\par\end{centering}
\caption{Fine structure constant $\alpha_{f}=e^{2}/(\epsilon v)$ vs lateral
spread of the wavefunctions in the quantum wires, $\ell$, normalized
by the interwire separation $L$. Curves show the boundary between
the Fermi liquid and smectic metal phases. Black triangles and purple
circles: spinless particles for $\lambda/L=0.5$ and $1$, respectively.
Blue square and orange diamonds: spinfull case for $\lambda/L=0.5$
and 1, respectively. At finite lateral spread $\ell$, there is a
quantum phase transition separating the strong coupling smectic metal
phase from the Fermi liquid one at weak coupling. At $\ell\to0$,
the quantum critical point collapses, and the smectic metal phase
is always stable. }
\end{figure}

\subsection{\textit{\emph{Backscattering effects}}}

\textcolor{black}{Orthogonality between left and right modes eliminates
backscattering in the present model. It is worth remarking that in
a more general model, where there is backscattering, the phase diagram
will comprise of regions where CDW coupling is the most relevant one.
In lattice models with Dirac fermions, a finite but very small amount
of backscattering is expected \cite{Killi}. Backscattering is also
expected in tWTe$_{2}$ bilayers and} in layered van der Waals material
NbSi$_{0.45}$Te$_{2}$ \cite{Yang-1}, where a non-symmorphic symmetry
protects directional massless Dirac fermions that form equally spaced
1D channels in the bulk of the material, akin to stripes.

In the spinless case, it is well known that intra-wire backscattering
can be rewritten as forward scattering term, thereby amounting only
to a redefinition of the Luttinger parameters \cite{Giamarchi}. In
the spinfull case, intra-wire backscattering is known to renormalize
the Luttinger parameters and introduce an irrelevant perturbation
(for repulsive interactions) in the spin channel \cite{Giamarchi,Fradkin}.
Backscattering between quantum wires, however, can be relevant and
may open a CDW gap at zero temperature. The coupling between wires
has the form
\begin{equation}
\mathcal{H}_{cdw}=g_{cdw}\sum_{a,\alpha,\sigma,\sigma'}\int dx\hat{\zeta}_{\alpha,\sigma}^{\dagger a}\hat{\zeta}_{-\alpha,\sigma}^{a}\hat{\zeta}_{-\alpha,\sigma'}^{\dagger a+1}\hat{\zeta}_{\alpha,\sigma'}^{a+1},\label{eq:cdw}
\end{equation}
and can lead to a CDW state. From the lowest order RG analysis, this
operator becomes relevant when $\eta_{cdw}=2-\triangle_{cdw}>0.$
For the spinless case, this is given by
\begin{equation}
\triangle_{cdw}=\int_{-\pi}^{\pi}\frac{dk_{\bot}}{2\pi}\left[2\kappa\left(k_{\bot}\right)^{-1}\right]\left(1-\cos k_{\bot}\right)
\end{equation}
where $\kappa\left(k_{\bot}\right)$ is defined in Eq. (\ref{kappa}).
For the spinful case, $\triangle_{cdw}^{\text{spin}}=1+\frac{1}{2}\triangle_{cdw}$. 

The plot in Fig. 3 shows the regions in the zero temperature phase
diagram where the operators corresponding to CDW, SC and Fermi liquid
phases become relevant, as a function of the fine structure constant
$\alpha_{f}$ and the ratio $\ell/L.$ In the model considered here,
$g_{cdw}=0$ due to orthogonality between left and right modes. In
lattice models that can be approximated by the continuum model discussed
before, $g_{cdw}$ can be non-zero, although still small. In this
scenario, the smectic metal phase may now give way for a more relevant
CDW state at zero temperature.

In models where the backscattering term $g_{cdw}$ is parametrically
small compared to forward scattering terms $g_{2}$ and $g_{4}$ to
begin with, the CDW gap is only observable at very small temperature.
In the RG spirit, temperature plays a role of an infrared cut-off
, where the RG flow stops. Since $g_{cdw}$ is a marginal operator
at the tree level ($g_{2}=g_{4}=0$), it grows under the RG only logarithmically
under reescaling of the momenta and fields, 
\begin{equation}
g_{cdw}(T)=g_{cdw}(\Lambda_{T})+\eta_{cdw}\ln\left(\frac{\Lambda_{T}}{T}\right),\label{eq:g-1}
\end{equation}
where $\Lambda_{T}$ is some ultraviolet temperature cut-off, with
$g_{cdw}(\Lambda_{T})\ll g_{2},g_{4}$. Hence, $g_{cdw}$ becomes
dominant over forward scattering processes only near zero temperature,
somewhere in the limit where $T/\Lambda_{T}\to0$. There must be hence
a low temperature $T_{*}$ above which backscattering effects are
subdominant, favoring either a smectic metal or Fermi liquid phases,
even when backscattering is the most relevant perturbation. This seems
to be the case in tWTe$_{2}$, where a smectic metal phase was observed
down to $1.8$K \cite{Wang}. We predict that placing tWTe$_{2}$
on a dielectric substrate at fixed $T>T_{*}$ could destabilize the
smectic metal towards a Fermi liquid phase.

\begin{figure}[t]
\begin{centering}
\includegraphics[scale=0.38]{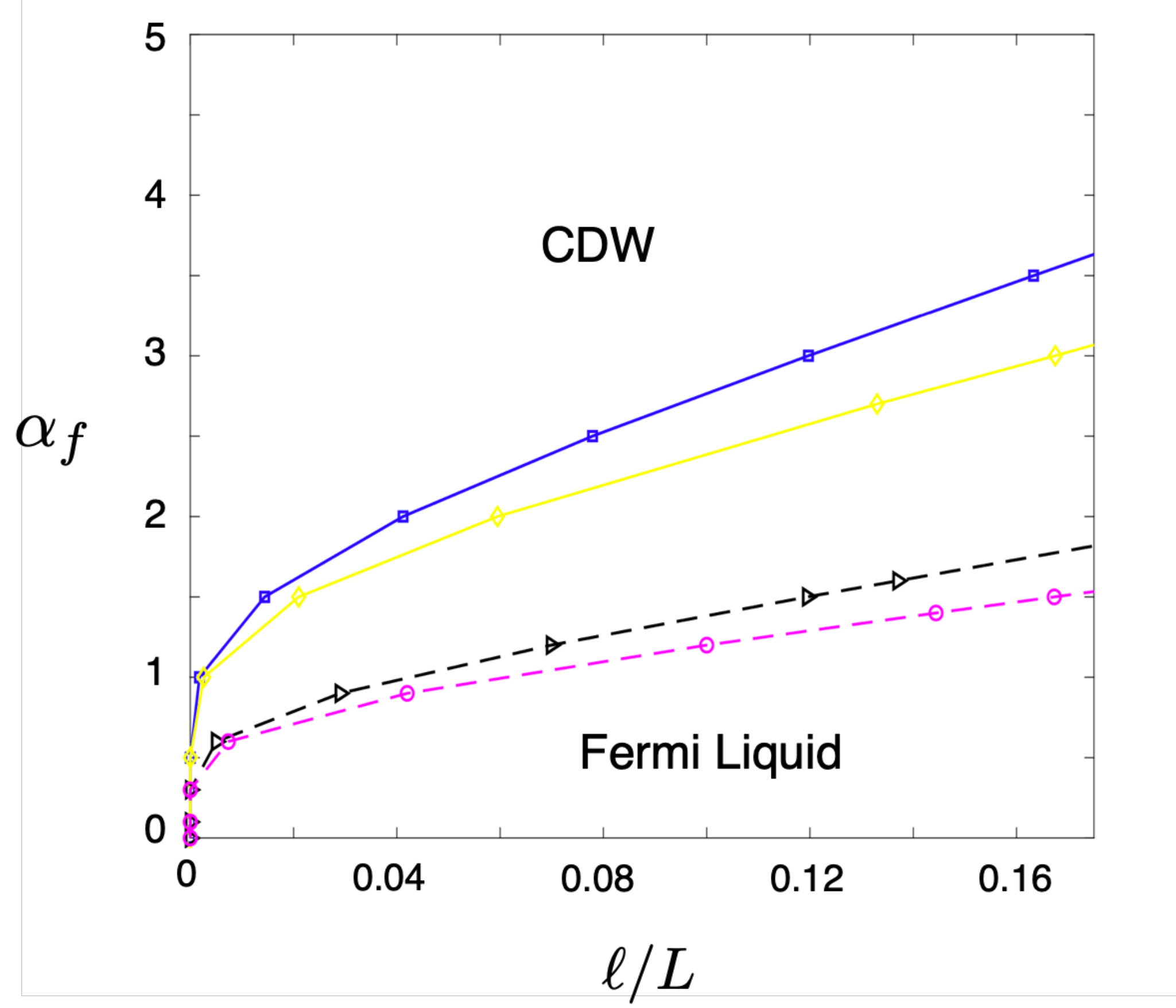}
\par\end{centering}
\caption{Zero temperature phase diagram in the presence of inter-wire backscattering.
Black triangles and magenta circles: spinless particles for $\lambda/L=0.5$
and 1, respectively. Blue square and yellow diamonds: spinful case
for $\lambda/L=0.5$ and 1, respectively. In models where inter-wire
backscattering is parametrically small compared to forward scattering,
the CDW phase can emerge only at very low temperature, even when the
most relevant perturbation to the smectic metal fixed point (see text).}
\end{figure}

The properties of those two phases, smectic metal and Fermi liquid,
are rather well known. While the Fermi liquid state is the most general
many particle state in two or three dimensions, the smectic metal
state is a rather peculiar state of matter. In a smectic metal, there
is large longitudinal conductivity in each quantum wire, but transport
is incoherent in the transverse direction due to the irrelevance of
inter-wire hopping. In the absence of disorder, resistivity along
the wires $\rho_{yy}=0.$ Small amounts of disorder, which are present
in realistic samples, can introduce backscattering and lead to a temperature
dependence $\rho_{yy}\sim T^{\alpha_{\Vert}},$ where $\alpha_{\Vert}=\left[\int_{-\pi}^{\pi}\frac{dk_{\bot}}{2\pi}\kappa\left(k_{\bot}\right)^{-1}\right]-2$
\cite{Emery,Mukhopadhyay,Luther}. On the other hand, in the transverse
direction, conductivity is still given by a power law, $\sigma_{xx}\sim T^{\alpha_{\bot}},$
where the exponent $\alpha_{\bot}$ depends on the details of single
particle hopping and Josephson couplings \cite{Mukhopadhyay}.

A few comments on the approximations used in our approach are in order.
The continuum Dirac model employed here is known to be a very good
approximation for the tight binding model of graphene at low energies.
We focused on the lowest energy modes that arise in this model in
the presence of a periodic mass term. In order to bosonize this model,
we project the Coulomb interactions onto these modes. Previous studies
have pointed out the deficiencies of the projected model, where processes
that are ignored may become important at intermediate energy scales
\cite{Meden}. To the least, we expect the field theoretical approach
used here to be a valid description of the asymptotic low-energy physics.

\section{Conclusion}

We considered an \textcolor{black}{effective} model for an array of
parallel quantum wires in 2D that accounts for the lateral spread
of the wavefunctions $\ell$ in the transverse direction to the wires.
The model lacks backscattering, and does not lead to a stripe phase.
Weak backscattering effects are expected to reintroduce a stripe phase
at very low temperature, which we discuss in detail. Using standard
abelian bosonization and RG methods, we calculated the Luttinger parameters
of the sliding LL phase in terms of effective parameters of the Hamiltonian
and analyzed what known instabilities of the smectic fixed point (previously
found phenomenologically) actually survive. We showed that the smectic
metal phase is stable in the ideal quantum wire limit $(\ell\to0)$,
and survives at finite $\ell$ beyond a critical Coulomb coupling
$\alpha_{c}(\ell)$ that grows monotonically with $\ell$. In weak
coupling ($\alpha<\alpha_{c}$), this model describes a 2D Fermi liquid,
with the wavefunctions in the quantum wires percolating over the whole
system. We find that superconductivity is absent, a feature that is
expected to be generic of similar models with Coulomb interactions.

\section{Acknowledgements }

GJ and BU thank Carl T. Bush Fellowship for support. BU thanks NSF
grant DMR-2024864 for partial support.


\begin{thebibliography}{10}
\bibitem{Giamarchi}T. Giamarchi, \textit{Quantum Physics in One Dimension}
(Oxford University Press, 2003).

\bibitem{Fradkin}E. Fradkin, \textit{Field Theories of Condensed
Matter Physics}, 2nd ed. (Cambridge University Press, 2013).

\bibitem{Strong}S. Strong, D. G. Clarke, and P. W. Anderson, Magnetic
Field Induced Confinement in Strongly Correlated Anisotropic Materials,
Phys. Rev. Lett. \textbf{73}, 1007 (1994).

\bibitem{Anderson}P. W. Anderson, \textquoteleft \textquoteleft Confinement\textquoteright \textquoteright{}
in the one-dimensional Hubbard model: Irrelevance of single-particle
hopping, Phys. Rev. Lett. \textbf{67}, 3844 (1991).

\bibitem{Tranquanda}J. M. Tranquada, D. J. Buttrey, V. Sachan, and
J. E. Lorenzo, Simultaneous Ordering of Holes and Spins in $\mathrm{La_{2}NiO_{4.125}}$,
Phys. Rev. Lett. \textbf{73}, 1003 (1994).

\bibitem{Zannen}J. Zaanen and O. Gunnarsson, Charged magnetic domain
lines and the magnetism of high- $T_{c}$ oxides, Phys. Rev. B \textbf{40},
7391 (1989).

\bibitem{Machida}K. Machida, Magnetism in $\mathrm{La_{2}CuO_{4}}$
based compounds, Phys. C Supercond. \textbf{158}, 192 (1989).

\bibitem{Kato}M. Kato, K. Machida, H. Nakanishi, and M. Fujita, Soliton
Lattice Modulation of Incommensurate Spin Density Wave in Two Dimensional
Hubbard Model -A Mean Field Study- , J. Phys. Soc. Japan \textbf{59},
1047 (1990).

\bibitem{Emery-3}V. J. Emery, S. A. Kivelson, and J. M. Tranquada,
Stripe phases in high-temperature superconductors, Proc. Natl. Acad.
Sci. USA \textbf{96}, 8814 (1999).

\bibitem{Mukhopadhyay-1}R. Mukhopadhyay, C. L. Kane, and T. C. Lubensky,
Crossed sliding Luttinger liquid phase, Phys. Rev. B \textbf{63},
081103(R) (2001).

\bibitem{Emery-1}See V. Emery, in Highly Conducting One-Dimensional
Solids, edited by J. Devreese et al. (Plenum, New York, 1979).

\bibitem{Bourbonnais}C. Bourbonnais and L. G. Caron, Renormalization
Group Approach To Quasi-One-Dimensional Conductors, Int. J. Mod. Phys.
B \textbf{5}, 1033 (1991).

\bibitem{Emery}V. J. Emery, E. Fradkin, S. A. Kivelson, and T. C.
Lubensky, Quantum Theory of the Smectic Metal State in Stripe Phases,
Phys. Rev. Lett. \textbf{85}, 2160 (2000).

\bibitem{Vishwanath}A. Vishwanath and D. Carpentier, Two-Dimensional
Anisotropic Non-Fermi-Liquid Phase of Coupled Luttinger Liquids, Phys.
Rev. Lett. \textbf{86}, 676 (2001).

\bibitem{Mukhopadhyay} R. Mukhopadhyay, C. L. Kane, and T. C. Lubensky,
Sliding Luttinger liquid phases, Phys. Rev. B \textbf{64}, 045120
(2001).

\bibitem{Chen}C. Chen, A. H. Castro Neto, and V. M. Pereira, Correlated
states of a triangular net of coupled quantum wires: Implications
for the phase diagram of marginally twisted bilayer graphene, Phys.
Rev. B \textbf{101}, 165431 (2020).

\bibitem{Xu}S. G. Xu \emph{et al.,} Giant oscillations in a triangular
network of one-dimensional states in marginally twisted graphene,
Nat. Comm. \textbf{10}, 4008 (2019).

\bibitem{Andrei}E. Y. Andrei, D. Efetov, P. Jarillo-Herrero, A. H.
MacDonald, K. F. Mak, T. Senthil, E. Tutuc, A. Yazdani, and A. F.
Young, The marvels of moiré materials, Nat. Rev. Mater \textbf{6},
201 (2021).

\bibitem{Wang}P. Wang, G. Yu,Y. H. Kwan, \emph{et al.,} One-dimensional
Luttinger liquids in a two-dimensional moiré lattice, Nature \textbf{605},
57--62 (2022).

\bibitem{Kindermann-1}M. Kindermann and P. N. First, Local sublattice-symmetry
breaking in rotationally faulted multilayer graphene, Phys. Rev. B
\textbf{83}, 045425 (2011).

\bibitem{Uchoa}B. Uchoa, V. N. Kotov, and M. Kindermann, Valley order
and loop currents in graphene on hexagonal boron nitride, Phys. Rev.
B \textbf{91}, 121412(R) (2015).

\bibitem{Yankowitz}M. Yankowitz, J. Xue, D. Cormode, J. D. Sanchez-Yamagishi,
K. Watanabe, T. Taniguchi, P. Jarillo-Herrero, P. Jacquod, and B.
J. LeRoy, Emergence of superlattice Dirac points in graphene on hexagonal
boron nitride, Nat. Phys. \textbf{8}, 382 (2012).

\bibitem{Yang}W. Yang, G. Chen, Z. Shi, C.-C. Liu, L. Zhang, G. Xie,
M. Cheng, D. Wang, R. Yang, D. Shi, K. Watanabe, T. Taniguchi, Y.
Yao, Y. Zhang, and G. Zhang, Epitaxial growth of single-domain graphene
on hexagonal boron nitride, Nat. Mater. \textbf{12}, 792 (2013).

\bibitem{Sachs}B. Sachs, T. O. Wehling, M. I. Katsnelson, and A.
I. Lichtenstein, Adhesion and electronic structure of graphene on
hexagonal boron nitride substrates, Phys. Rev. B \textbf{84}, 195414
(2011).

\bibitem{Kindermann}M. Kindermann, B. Uchoa, and D. L. Miller, Zero-energy
modes and gate-tunable gap in graphene on hexagonal boron nitride,
Phys. Rev. B \textbf{86}, 115415 (2012).

\bibitem{Killi}M. Killi, Tzu-Chieh Wei, I. Affleck, and A. Paramekanti,
Tunable Luttinger Liquid Physics in Biased Bilayer Graphene, Phys.
Rev. Lett. \textbf{104}, 216406 (2010).

\bibitem{Martin}I. Martin, Y. M. Blanter, and A. F. Morpurgo, Topological
Confinement in Bilayer Graphene, Phys. Rev. Lett. \textbf{100}, 036804
(2008).

\bibitem{Biswas} S. Biswas, T. Mishra, S. Rao, and A. Kundu, Chiral
Luttinger liquids in graphene tuned by irradiation, Phys. Rev. B \textbf{102},
155428 (2020).

\bibitem{Heeger}A. J. Heeger \emph{et al.}, Solitons in conducting
polymers, Rev. Mod. Phys. \textbf{60}, 781 (1988).

\bibitem{Semenoff}G. W. Semenoff, V. Semenoff, and F. Zhou, Domain
Walls in Gapped Graphene, Phys. Rev. Lett. \textbf{101}, 087204 (2008).

\bibitem{Jackiw}R. Jackiw and C. Rebbi, Solitons with fermion number
$\sfrac{1}{2}$, Phys. Rev. D \textbf{13}, 3398 (1976).

\bibitem{Tchoumakov}S. Tchoumakov, V. Jouffrey, A. Inhofer, E. Bocquillon,
B. Plaçais, D. Carpentier, M. O. Goerbig, Volkov-Pankratov states
in topological heterojunctions, Phys. Rev. B \textbf{96}, 201302 (2017).

\bibitem{Castro-Neto}A. H. Castro Neto, F. Guinea, N. M. R. Peres,
K. S. Novoselov, and A. K. Geim, The electronic properties of graphene,
Rev. Mod. Phys. \textbf{81}, 109 (2009).

\bibitem{text-1}In comparison with the notation in \cite{Emery},
$\left(\frac{u^{\rho}(k_{\bot})}{K^{\rho}(k_{\bot})}\right)^{-1}=W_{0}(k_{\bot})$
and $u^{\rho}(k_{\bot})K^{\rho}(k_{\bot})=W_{1}(k_{\bot}).$ The parameter
that decides the phase diagram is $\kappa=\sqrt{W_{0}(k_{\bot})W_{1}(k_{\bot})}.$

\bibitem{note0}The Coulomb term can be cast in terms of charge and
current density terms $j_{\mu}(\mathbf{r})W_{\mu}(\mathbf{r}-\mathbf{r}^{\prime})j_{\mu}(\mathbf{r}^{\prime})$,
where $j_{0}=\rho_{+}+\text{\ensuremath{\rho}}_{-}$ is a charge density
and $j_{1}(\mathbf{r})=\rho_{+}-\text{\ensuremath{\rho}}_{-}$ the
current density. When $g_{2}=g_{4}$ the current-current term is zero.

\bibitem{text-2}Notice that, for the spinless case here there is
an extra factor of 2, compared to \cite{Emery}. This is because we
consider a spinless case as opposed to the spin gapped case considered
in high-$T_{c}$ like scenarios.

\bibitem{Text} One could in principle consider the Josephson coupling
between next-nearest neighbor and so on after including the same number
of terms in \eqref{Luttinger-parameters-smectic}. We have checked
that $sc$ coupling for the next nearest neighbor is even less relevant.

\bibitem{Emery-2}V. J. Emery, S. A. Kivelson, and O. Zachar, Spin-gap
proximity effect mechanism of high-temperature superconductivity,
Phys. Rev. B \textbf{56}, 6120 (1997).

\bibitem{Kim2}M. Kim, \emph{et al}., Control of electron-electron
interaction in graphene by proximity screening, Nat. Commun. \textbf{11},
2339 (2020).

\bibitem{Yang-1}T. Y. Yang, \emph{et. al., }Directional massless
Dirac fermions in a layered van der Waals material with one-dimensional
long-range order, Nat. Materials \textbf{19}, 27 (2020).

\bibitem{key-2}Umklapp interactions are relevant for repulsive interactions
and can open a CDW gap (see Ref. ). These have a non-zero amplitude
only precisely at half filling and can be ignored in experimental
settings where the chemical potential can be tuned at will. 

\bibitem{Luther}A. Luther and I. Peschel, Fluctuation Conductivity
and Lattice Stability in One Dimension, Phys. Rev. Lett. \textbf{32},
922 (1974).

\bibitem{Pereira}V. M. Pereira, A. H. Castro Neto, and N. M. R. Peres,
Tight-binding approach to uniaxial strain in graphene, Phys. Rev.
B. \textbf{80}, 045401 (2009). 

\bibitem{Meden}Meden, V., Metzner, W., Schollwöck, U. et al. Luttinger
liquids with boundaries: Power-laws and energy scales, Eur. Phys.
J. B \textbf{16}, 631--646 (2000)
\end{thebibliography}
\end{document}